\documentclass[11pt,twocolumn]{article}
\usepackage{times}
\usepackage{mathfont}
\usepackage{url}

\def\square{\framebox{ } \smallskip \smallskip}
\long\def\omit#1{} 
\def\log{\mathop{{\rm log}}}
\def\Pr{\mathop{{\rm Pr}}}

\mathcode`O="724F

\newtheorem{theorem}{Theorem}
\newtheorem{lemma}[theorem]{Lemma}

\begin{document}
\title{Fast Approximation of Centrality}
\author{David Eppstein\thanks{Dept. Inf. \& Comp. Sci., 
UC Irvine, CA 92697-3425, USA,
{\tt\{eppstein,josephw\}@ics.uci.edu}.}
\and Joseph Wang$^*$}
\date{ }
\maketitle

\begin{abstract}
Social studies researchers use graphs to model
group activities in social networks.
An important property in this context is
the {\em centrality} of a vertex: the inverse of the
average distance to each other vertex.  We describe a
randomized approximation algorithm for centrality in weighted
graphs.  For graphs exhibiting the small world phenomenon, our
method estimates the centrality of all vertices
with high probability within a $(1+\epsilon)$ factor in near-linear time.
\end{abstract}

\section{Introduction}
In social network analysis, the vertices of a graph represent 
agents in a group and the edges represent relationships, such
as communication or friendship.
The idea of applying graph theory to analyze the
connection between the structural {\em centrality} 
and group process was introduced by Bavelas \cite{Bavelas48}. 
Various measurement of centrality \cite{Bonacich72,Freeman79,Friedkin91}
have been proposed for analyzing
communication activity, control, or independence within 
a social network.

We are particularly interested in {\em closeness centrality}
\cite{Bavelas50,Beauchamp65,Sabidussi66}, which is used to
measure the independence and efficiency of an agent 
\cite{Freeman79,Friedkin91}. Beauchamp~\cite{Beauchamp65} defined 
the closeness centrality of agent $a_j$ as
$${n - 1} \over {\sum_{i = 1}^{n} d(i, j)}$$
where $d(i, j)$ is the
distance between agents $i$ and~$j$.\footnote{This
should be distinguished from another common concept of graph centrality,
in which the most central vertices minimize the maximum
distance to another vertex.}
We
are interested in computing centrality  values for all agents.
To compute the centrality for each agent, 
it is sufficient to solve the all-pairs shortest-paths (APSP) 
problem. No faster exact method is known.

The APSP problem can be solved by various algorithms
in time $O(nm + n^2 \log n)$ \cite{FredmanTarjan87,Johnson77}, $O(n^3)$
\cite{Floyd62}, or more quickly using fast matrix multiplication
techniques \cite{AGM97,CoppWin90,Seidel95,Yuval76}.
\omit{Several researchers have developed more efficient algorithms for
special graph classes such as interval graphs 
\cite{ACL93,CLSS98,RMPR92} and chordal graphs \cite{BCD94,HSS97}.
The APSP problem
can be solved in average-case in time $O(n^2 \log n)$
for various classes of random graphs  
\cite{CFMP97,FriezeGrimmett85,MehlhornPriebe95,MoffatTakaoka85}.}
Because these results are slow
or (with fast matrix multiplication) complicated and impractical,
and because recent applications of social network theory to the internet
may involve graphs with millions of vertices, it is of interest to
consider faster approximations. Aingworth et al. \cite{ACIM99} proposed 
an algorithm with an additive error of $2$
for the unweighted APSP problem
that runs in time $O(n^{2.5}\sqrt{\log n})$.
However this is still slow and does not provide a good approximation
when the distances are small.

In this paper, we consider a method for fast approximation of centrality.  
We apply a random sampling technique to approximate the 
inverse centrality of all vertices in a weighted graph to within an
additive error of $\epsilon \Delta$ with high probability
in time $O({\log n \over \epsilon^2} (n \log n + m))$, where
$\epsilon$ is any fixed constant and $\Delta$ is the diameter of the
graph.

It has been observed empirically that many social networks exhibit the
{\em small world phenomenon} \cite{Milgram67}: their diameter is bounded
by a constant, or, equivalently, the ratio between the minimum and
maximum distance is bounded.  For such networks, the inverse centrality
at any vertex is $\Omega(\Delta)$ and our method provides a near-linear
time $(1+\epsilon)$-approximation to the centrality of all vertices.

\omit{\section{Preliminaries} 
We are given a graph $G(V, E)$ with $n$ vertices
and $m$ edges. The distance $d(u, v)$ between
two vertices $u$ and $v$ is the length of the shortest path
between them. The diameter $\Delta$ of a graph $G$
is defined as $max_{u, v \in V} d(u, v)$. For
simplicity, we define centrality $c_v$ for vertex $v$
as ${n - 1} \over {\sum_{u \in V} d(u, v)}$. If $G$ is not
connected, then $c_v = \infty$. Hence we will assume $G$ is connected.}

\omit{Given an optimization problem $P$. Let $value(OPT)$ denote
the optimal solution for a problem instance in $P$.
Let $value(A)$ denote the solution computed by
an approximation algorithm $A$.
$A$ is said to have constant additive approximation
error $c$ if $|value(A) -  value(OPT)| \le c$ for every 
problem instance in $P$. }

\section{The Algorithm}

We now describe a randomized 
approximation algorithm RAND for estimating centrality.
RAND randomly chooses $k$ sample vertices and computes
single-source shortest-paths (SSSP) from each sample vertex to all
other vertices. The estimated centrality of a vertex is
defined in terms of the average distance to the sample vertices.

\vfil\eject

\noindent {\bf Algorithm RAND:}
\begin{enumerate}
\item Let $k$ be the number of iterations needed to
obtain the desired error bound. 
\item In iteration $i$, pick 
vertex $v_i$ uniformly at random from $G$ and solve the SSSP problem
with
$v_i$ as the source.
\item Let
	$$\hat{c}_u = 1/\sum_{i = 1}^{k}\frac{n\,d(v_i, u)}{k(n-1)}$$
be the centrality estimator for vertex $u$.
\end{enumerate}

\smallskip
It is not hard to see that, for any $k$ and $u$,
the expected value of $1/\hat{c}_u$ is equal to $1/c_u$.  

\omit{
PROOF NEEDS FIXING TO ACCOUNT FOR N/(N-1) FACTOR!
\begin{theorem}
$E[1/\hat{c}_u] = 1/c_u$.
\end{theorem}

{\bf Proof:} 
Each vertex has equal probability of $1/n$ to be picked at each
round. The expected value for $1 \over \hat{c}_u$ is
\begin{eqnarray*}
E[1 \over {\hat{c}_u}] & = & {n \over n - 1} 1/n^{k} \cdot {{kn^{k - 1}
\sum_{i = 1}^{n} d(i, u)} \over k} \\
& = & {n \over n - 1} {{\sum_{i = 1}^{n} d(i, u)} \over n}  \\
& = & {1 \over c_u}.\end{eqnarray*}         
\square
\bigskip
}

\omit{In 1963, Hoeffding \cite{Hoeffding63} gave the following theorem 
on probability bounds for sums of independent random variables.}

\begin{lemma}[Hoeffding 
\cite{Hoeffding63}]
If $x_1, x_2, \ldots, x_{k}$ are independent,
$a_i \le x_i \le b_i$,
and $\mu = E[\sum x_i/k]$ is the expected mean,
then for $\xi > 0$
$$\Pr\Bigl\{ |{\sum_{i = 1}^{k} x_i \over k} - \mu| \ge \xi \Bigr\}
\le 2 e^{-2{k}^2 {\xi}^2/\sum_{i = 1}^{k}(b_i - a_i)^2}.$$
\end{lemma}
\bigskip

We need to bound the probability that the error in estimating
the inverse centrality of any vertex $u$ is at most $\xi$.
This is done by applying Hoeffding's bound with
$x_i = \frac{d(i, u) n}{(n-1)}$,
$\mu = \frac{1}{c_u}$,
$a_i=0$, and $b_i=\frac{n\Delta}{n-1}$. 
\omit{
We know $E[1/\hat{c}_u] = 1/c_u$.
To take care of the case in
which $\hat{c}_u$ is smaller than $c_u$, we multiply
the above inequality by $2$.
}
Thus the probability that 
the difference between the estimated inverse centrality
$1/\hat{c}_u$ and the actual inverse centrality $1/c_u$ is more than $\xi$
is
\begin{eqnarray*}
\Pr\left\{ {\textstyle |\frac{1}{\hat{c}_u} - \frac{1}{c_u}|}
\ge \xi \right\}  
& \le &
2 \cdot e^{-2{k}^2 {\xi}^2/\sum_{i = 1}^{k}(b_i - a_i)^2} \\
& = & 2 \cdot e^{-2{k}^2 {\xi}^2/{k}(\frac{n\Delta}{n-1})^2} \\
& = & 2 \cdot e^{-\Omega(k\xi^2/\Delta^2)}
\end{eqnarray*}
For $\xi = \epsilon\Delta$, using $\Theta(\frac{\log n}{\epsilon^2})$
samples will cause the probability of error at any vertex to be bounded
above by e.g. $1/n^2$, giving at most $1/n$ probability of
having greater than $\epsilon\Delta$ error anywhere in the graph.

\omit{
Fredman and Tarjan \cite{FredmanTarjan87} gave an
algorithm for solving the $SSSP$ problem in time $O(n \log n + m)$.
Thus }
The total running time of algorithm is
$O(k \cdot m)$ for unweighted graphs and $O(k (n \log n + m))$
for weighted graphs.
Thus, for $k = \Theta(\frac{\log n}{\epsilon^2})$,
we have an $O({\log n \over \epsilon^2} (n \log n + m))$ algorithm
for approximating centrality within an inverse additive
error of $\epsilon \Delta$ with high probability.

\omit{\section{Conclusion}
We gave an $O({\log n \over \epsilon^2} (n \log n + m))$ 
randomized algorithm with additive error of $\epsilon \Delta$
for weighted graphs. Many graph classes such as paths, cycles,
and balanced trees, have centrality proportional to
$\Delta$. More interestingly, Milgram \cite{Milgram67} showed that 
many social networks have bounded diameter and centrality.
When the centrality is proportional to $\Delta$,
we have an $(1 + \epsilon)$-approximation algorithm. }

\small
\paragraph{Acknowledgements.}
We thank Dave Goggin for bringing this problem to our attention,
and Lin Freeman for helpful comments on a draft of this paper.

\bibliographystyle{nomonths}
\let\oldbib\thebibliography
\def\thebibliography#1{\oldbib{#1}\itemsep 0pt}
\bibliography{bibdata}
\end{document}